\documentclass[reqno,10pt]{article}
\usepackage{graphicx}
\usepackage{amsmath}
\usepackage{amssymb}
\usepackage{amsthm}
\usepackage{enumitem}
\usepackage{bbm}
\usepackage[l2tabu,orthodox]{nag}
\usepackage[all,error]{onlyamsmath}
\usepackage[strict]{csquotes}
\usepackage{url}
\newtheorem{theorem}{Theorem}

\theoremstyle{definition}

\numberwithin{equation}{section}
\numberwithin{theorem}{section}

\newenvironment{OMabstract}{\noindent\textbf{Abstract.} }{\medskip}
\newenvironment{OMsubjclass}{\noindent\textbf{Mathematics Subject Classification (2020):} }{\medskip}
\newenvironment{OMkeywords}{\noindent\textbf{Keywords:}  }{\medskip}

\begin{document}

\author{Vyacheslav Pivovarchik} 
\title{ Recovering the shape of a quantum tree by scattering data}
\maketitle


\begin{OMabstract}
     We consider a scattering problem generated by the Sturm-Liouville equation on a tree which consists of a equilateral compact subtree with a lead (a half-infinite edge) attached to this compact subtree.  We assume that the potential on the lead is zero identically and the potentials on the finite edges are $L_2$-functions.  
We show how to find the shape of the   tree using the S-function of the scattering problem.  
\end{OMabstract}

\begin{OMkeywords}
      Sturm-Liouville equation, eigenvalue, equilateral tree, star graph, Dirichlet boundary condition, Neumann boundary condition, lead, S-function, asymptotics.
\end{OMkeywords}

\begin{OMsubjclass}
    34B45, 34B240, 34L20
\end{OMsubjclass}


\renewcommand{\baselinestretch}{1.5}
\newcommand{\C}{{\bf C}}
\newcommand{\CB}{{\cal B}}
\newcommand{\CH}{{\cal H}}
\newcommand{\CL}{{\cal L}}
\newcommand{\cotanh}{\mbox{cotanh}}
\newcommand{\HB}{{\cal HB}}
\newcommand{\N}{{\bf N}}
\newcommand{\R}{{\bf R}}
\newcommand{\sgn}{\mbox{sgn}}
\newcommand{\SHB}{{\cal SHB}}
\newcommand{\Z}{{\bf Z}}
\newcommand{\za}{\alpha}
\newcommand{\zb}{\beta}
\newcommand{\ze}{\varepsilon}
\newcommand{\zf}{\varphi}
\newcommand{\zg}{\gamma}
\newcommand{\zk}{\kappa}
\newcommand{\zl}{\lambda}
\newcommand{\zo}{\omega}
\newcommand{\zs}{\sigma}
\newcommand{\zz}{\zeta}
\newcommand{\rf}[1]{(\ref{#1})}
\newcommand{\p}{^{\prime}}
\newcommand{\pp}{^{\prime\prime}}
%
%


\section{Introduction.}
\setcounter{equation}{0}

We consider the Sturm-Liouville scattering problem and the problem of recovering the shape of a metric graph  consisting of a compact tree and a lead (half-infinite edge) attached to it. As far as we know, the first result  on this problem was obtained in \cite{GS} where it was proved that if the lengths of the edges are non-commensurate then the S-function uniquely determines the shape of the graph. In general, the knowledge of the $S$-matrix is not sufficient to determine the
topological structure of the graph uniquely: several negative results have been obtained in~\cite{KurSte02}.

A substantial (first?) attempt to tackle the question of taking cospectral objects and checking whether adding scattering data helps to distinguish them was made by Okada, Shudo et al. in 2005 in \cite{OSh}
In particular, they took the very classical example of Gordon and Webb and checked its scattering properties.
Motivated by this work R. Band, A. Sawicki and U. Smilansky checked what happens for metric graphs \cite{BSS}.
They  found that scattering might not distinguish between cospectral graphs, if the leads are connected in a special way (which somehow resembles the symmetry behind the isospectral construction).
In \cite{BSS1} the authors explained why there is a difference between the results of the Japanese group (that scattering data resolves cospectrality) and between their results (that it does not). After these theoretical works, A. Sawicki continued collaborating with Polish experimental physicists who have actually built such microwave networks and examined the problem there.  See for example in \cite{HLBSKS}.

Here we continue investigation started in \cite{MuP} where it was shown that if we attach a lead to a compact simple equilateral graph then the S-function together with the eigenvalues uniquely determine the shape of the graph if the number of vertices is $\leq 6$ and if he compact subgraph is an equilateral tree  then this statement is true for the number of vertices is $\leq 9$.

In present paper we show how to construct the shape of a tree using the  S-function. As in \cite{MuP} we assume that the potential on the lead is identically 0 to deal with a meromorphic S-function. This approach originates from \cite{Re} and was used to deal with quantum graphs in \cite{P1} and \cite{LP}. In this case  S-function is meromorphic and the Jost function which is an entire function of exponential type.

In Section 2 we consider some results on combinatorial trees obtained in \cite{P}.  Namely, we show that  the ratio of the characteristic polynomial (the determinant of the normalized Laplacian) to the determinant of the modified  normalized Laplacian of the graph (forest) obtained by deleting the root of the tree together with the incident edges can be expanded  into a branched continued fraction of special form. The coefficients in this continued fraction are the degrees of the vertices of the initial tree. 

In Section 3 we describe spectral problems generated by the Sturm-Liouville equation on an equilateral tree. We consider the Neumann problem, i.e. spectral problem with standard conditions (the continuity and Kirchhoff's conditions at the interior vertices and the Neumann conditions at the pendant vertices). By Dirichlet problem we mean the problem with the Dirichlet condition at the root of a tree and standard conditions at the rest of the vertices.

In Section 4 we describe a scattering problem on a noncompact tree consisting of a lead (half-infinite edge) attached to a compact metric equilateral subtree. The corresponding operator is self-adjoint. The essential spectrum of this operator covers the non-negative half-axis. Also there can exist normal eigenvalues (isolated eigenvalues of finite multiplicity) and eigenvalues embedded into the essential spectrum.

In Section 5 we show how to find the shape of a tree using the scattering S-function.

\section{Auxiliary results}

In this section we consider combinatorial trees and forests.

Let ${\cal T}$ be a  combinatorial tree of $p$ vertices rooted at $v_0$ and let $A$ be its adjacency matrix with the first row corresponding to $v_0$. Let
\begin{equation}
\label{2.1}
 D=diag\{d(v_0), d(v_1),...,d(v_{p-1})\}
\end{equation} 
be the diagonal degree matrix (by $d(v)$ we denote the degree of vertex $v$).   Let $\hat{A}$ be the principal submatrix of $A$ obtained by deleting the first row and the first column from $A$ and $\hat{D}$ be the diagonal submatrix obtained by deleting the first row and the first column from $D$. Denote by
\begin{equation}
\label{2.2}
\psi(z):=det(-zD+A)
\end{equation}
the nomalized Laplacian of ${\cal T}$ and by 
\[\hat{\psi}(z):=det(-z\hat{D}+\hat{A}).
\]
This polynomial can be called modified normalized Laplacian of a tree or a forest $\hat{{\cal T}}$ obtained by deleting the root together with its incident edges. Modified, because  the entries in $\hat{D}$ are the degrees of the vertices in ${\cal T}$ (not in ${\hat{\cal T}}$).

The following theorem was proved in \cite{P} (Theorem 3.1).

{\bf Theorem 2.1}
{\it Let ${\cal T}$ be a tree. Then the fraction $\frac{\psi(z)}{\hat{\psi}(z)}$ can be  expanded in branched continuous fraction in such a way that the coefficients before $+z$ or $-z$ correspond to the degrees of  the vertices. The beginning fragment 
\[
-m_0z+
\sum_{k=1}^{m_0} \frac{1}{
m_kz-...}
\]
of the expansion means that the root $v_0$ is connected by edges with  $m_0$ vertices, say $v_1$, $v_2$, ..., $v_{m_0}$.

A fragment
\[
...\pm\sum_{i=1}^{r}\frac{1}{-m_iz+\sum_{k=1}^{m_i-1}\frac{1}{+m_{i,k}z+...}}
\]
means that there are $r$ vertices each have one incoming edge and $m_i-1$ ($i=1,...,r$)
 outgoing edges.

A fragment 
\[
... +\frac{m}{z}
\]
at an end of a branch of the continuous fraction means $m$ edges ending with  pendant vertices.}

\vspace{3mm}


By snowflake graph we mean a tree consisting of star graphs joined in the form shown at Fig. 1. The following result was obtained in \cite{P} (Theorem 3.2).

\vspace{3mm}


\begin{figure}
\begin{center}
   \includegraphics [scale= 0.7 ] {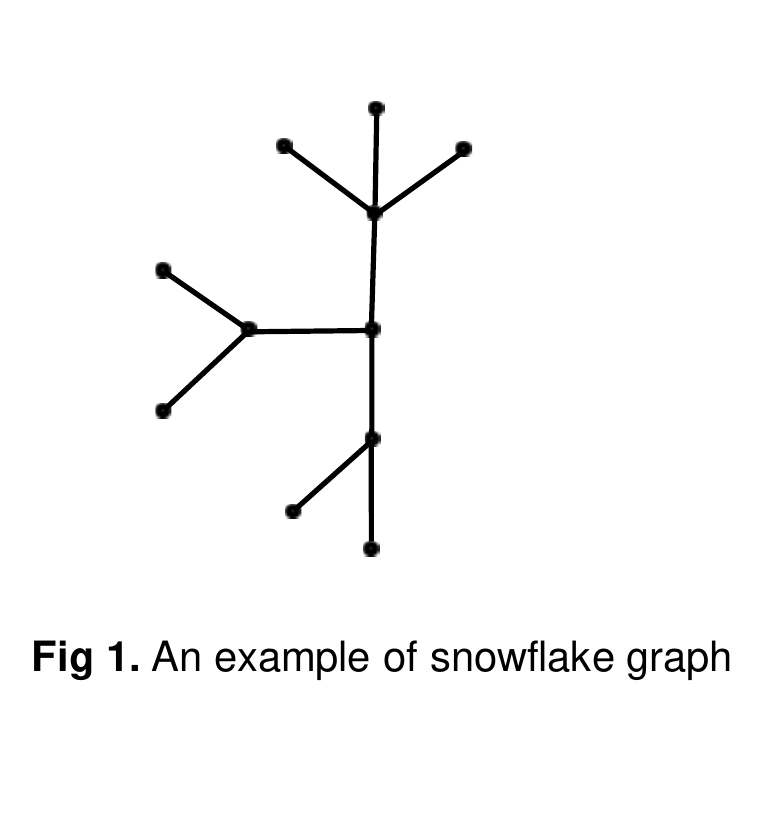}
\end{center}
\end{figure}

{\bf Theorem 2.2} {\it  Let ${\cal T}$ be a snowflake graph rooted at the central vertex. The corresponding  two functions  $\psi(z)$ and $\hat{\psi}(z)$ uniquely determine the shape of the graph. }

In case of snowflake graph we have
\[
\frac{\psi(z)}{\hat{\psi}(z)}=
-d(v_0)z+
\sum_{k=1}^{d(v_0)} \frac{1}{
d(v_k)z-\frac{d(v_k)-1}{z}}
\]
where $d(v_0)$ is the degree of the the root, $d(v_k)$ ($k=1,2, ..., d(v_0)$) are the degrees of the other interior vertices.

\section{Quantum graph problems}

Let $T$ be an equilateral metric  tree with $p$ vertices and $g=p-1$ edges each of the length $l$. We choose an arbitrary  vertex $v_0$ as the root and direct all  the edges  away from the root.  
Let us describe the {\it Neumann} spectral  problem on this tree. We consider  the Sturm-Liouville equations on the edges 
\begin{equation}
\label{3.1}
-y_j^{\prime\prime}+q_j(x)y_j=\lambda y_j, \ \  j=1,2,..., g 
\end{equation} 
where $q_j\in L_2(0,l)$ are real.

For each edge $e_j$ incident with a pendant vertex which is not the root 
we impose the Neumann condition 
\begin{equation}
\label{3.2}
y_j'(l)=0. 
\end{equation}
At  each interior vertex which is not the root we impose the continuity conditions   
\begin{equation}
\label{3.3}
y_j(l)=y_k(0)
\end{equation}
for the incoming to $v_i$ edge $e_j$ and for all $e_k$ outgoing from $v_i$ and the Kirchhoff's conditions
\begin{equation}
\label{3.4}
y'_j(l)=\mathop{\sum}\limits_k y_k'(0)
\end{equation}
where the sum is taken over all the edges $e_k$ outgoing from $v_i$. 

If the root is an interior vertex then the conditions at $v_0$ are
\begin{equation}
\label{3.5}
y_i(0)=y_j(0)
\end{equation}
for all indices $i$ and $j$ of the edges incident with the root and
\begin{equation}
\label{3.6}
\mathop{\sum}\limits_i y_i'(0)=0.
\end{equation}
If the root is pendant then
\begin{equation}
\label{3.7}
y_1'(0)=0.
\end{equation}
The above conditions (continuity +Kirchhoff's or Neumann if the vertex is pendant)  we call standard.



{\bf Standing assumption}
For all edges the potentials $q_j$ are real-valued functions of the space $L^2(0,\ell)$.
\vspace{3mm}

In the sequel, if the potentials are the same on all the  edges we omit the index in $q_j$ and $y_j$.  

We associate the metric tree $T$ with a combinatorial tree ${\cal T}$ described in Section 2.

The following theorem is a version of Theorem 5.2 in \cite{CP} which is based on the results of \cite{vB} and \cite{CaP}.

{\bf Theorem  3.1} 
{\it
 Let $T$ be a tree with $p\geq 2$.  Assume that all edges have the same length $l$ and the same potentials symmetric with respect to the midpoints of the edges ($q(l-x)\mathop{=}\limits^{a.e.}q(x)$).  Then the spectrum of problem (\ref{3.1})--(\ref{3.6})  or (\ref{3.1})--(\ref{3.4}), (\ref{3.7}) coincides with the set of zeros of the function  
\begin{equation}
\label{3.8} 
\phi_N(\lambda)=s(\sqrt{\lambda}, l) \tilde{\psi}(c (\sqrt{\lambda}, l))
\end{equation} 
where $\tilde{\psi}(z)=(1-z^2)^{-1}\psi(z)$ and $\psi(z)$ is defined by (\ref{2.2}).
 
Here $A$ is the adjacenty matrix of $T$, 
$D$ is defined by (\ref{2.2}), $s(\sqrt{\lambda},x)$ and $c(\sqrt{\lambda},x)$ are the solutions of the Sturm-Liouville equation on the edges satisfying the  conditions $s(\sqrt{\lambda},0)=s'(\sqrt{\lambda},0)-1=0$ and $c(\sqrt{\lambda},0)-1=c'(\sqrt{\lambda},0)=0$.}

\vspace{4mm}

Now we consider the Dirichlet problem. We impose the Dirichlet condition at $v_0$: 
\begin{equation}
\label{3.9}
y_i(0)=0
\end{equation}
for all edges incident with $v_0$, and consider the Dirichlet problem which consists of equations (\ref{3.1})--(\ref{3.4}) and (\ref{3.9}).   

Then 
 we can consider  $T$ as a union of $d(v_0)$  subtrees   $T_1$, $T_2$, ..., $T _{d(v_0)}$ which have common vertex $v_0$ and spectral problems on them meaning that  the Dirichlet conditions are imposed at $v_0$ while at the rest of  vertices we keep the standard  conditions. Thus, we have $d(v_0)$ problems on the subtrees.  
 
 Denote by $\hat{T}_i$ the tree obtained by removing the pendant vertex with the Dirichlet boundary conditions (the root) and the edge  incident with it in $T_i$. 
 Let $\hat{A}_i$  be the adjacency matrix of ${\hat T}_i$, let
$\hat{D}_{T,i}=diag \{d(v_{i,1}), d(v_{i,2}), ..., d(v_{i,p_i-1})\}$, where 
$d(v_{i,j})$ is the degree of the vertex $v_{i,j}$ in $T_i$ and $p_i$ is the number of vertices $\{v_0, v_{i,1}, ..., v_{i,p_i-1}\}$ in $T_i$.

We consider the polynomial $\hat{\psi}_i(z)$ defined by 
\begin{equation}
\label{3.10}
\hat{\psi}_{i}(z):=det(z\hat{D}_{T,i}-\hat{A_i}).
\end{equation}
Theorem 6.4.2 of \cite{MP} adapted to the case a tree with the Dirichlet condition at one of the  vertices is as follows

{\bf Theorem 3.2} {\it Let $T_i$ be a subtree of $T_i$ with at least one edges rooted at a pendant vertex $v_0$. Let the Dirichlet condition be imposed at the root and the standard conditions at all other vertices. Assume that all edges have the same length $l$ and the same potentials symmetric with respect to the midpoints of the edges ($q(l-x)=q(x)$). Then the spectrum of problem (\ref{3.1})--(\ref{3.4}), (\ref{3.9})  coincides with the set of zeros of  the characteristic function
\begin{equation}
\label{3.11}
\phi_{D,i}(\lambda)=\hat{\psi}_i(c(\sqrt{\lambda},l)).
\end{equation}
 }


It is clear that 
\begin{equation}
\label{3.12}
\phi_D(\lambda)=\prod_{i=1}^{d(v_0)}\phi_{D,i}(\lambda)=\prod_{i=1}^{d(v_0)}det(c(\sqrt{\lambda},l)\hat{D}_{T,i}-\hat{A_i})
\end{equation} 
is the characteristic function of the Dirichlet problem  (\ref{3.1})--(\ref{3.4}), (\ref{3.9}) on the initial tree $T$.

 Denote by  
\begin{equation}
\label{**}
\hat{\psi}(z):=\prod_{i=1}^{d(v_0)}\hat{\psi}_{i}(z). 
\end{equation}

It is clear that
\[
\hat{\psi}(z)=det(-z\hat{D}+\hat{A}).
\]

\vspace{3mm}

\section{Attaching a lead to a compact metric trees}\label{sec:scattering}

In this section we consider a graph $T_\infty$ obtained by attaching a lead $e_0$ to $v_0$ which is the root of a compact equilateral tree $T$ described in the previous section. We direct this edge $e_0$ away from $v_0$ and assume that the potential $q_0$ on $e_0$ is identically $0$. Thus we have the equations
\begin{equation}
\label{4.1}
-y_i^{\prime\prime}(x)+q_i(x)y_e(x)=\lambda y_i(x),\qquad  i\in E ,\ x\in [0,\ell],
\end{equation}
on all finite edges along with the equation
\begin{equation}
\label{4.2}
-y_0^{\prime\prime}(x)=\lambda y_0(x),\qquad x\in [0,\infty),
\end{equation}
on the edge $e_0$.
We endow the Sturm--Liouville equation~\eqref{4.1}--\eqref{4.2} on $T_\infty$ with standard conditions at all vertices.

For a closed linear operator $B$ on a Hilbert space, we let $D(B)$, $\rho(B)$
and $\sigma(B)$ denote its domain, resolvent set and spectrum. We
refer to \cite[Section I.2]{GK} for the definition of \textit{normal} (that is,
isolated eigenvalues of finite multiplicity) eigenvalues, and denote by $\sigma_{0}(B)$ the
set of normal eigenvalues of $B$ and by
$\sigma_{\mathrm{ess}}(B)=\sigma(B)\backslash\sigma_{0}(B)$ the essential
spectrum. 
At this
point we recall that the spectrum of any self-adjoint operator $B$
coincides with its approximative spectrum, see, e.g.,
\cite[page 118]{EE}, where the latter is defined as the set of
$\lambda\in\mathbb{C}$ such that there exists a sequence
$\left\{f_{n}\right\}_{n=1}^\infty$ in $D(B)$
such that $\| f_{n}\|\equiv 1$ and $(\lambda I-B)f_{n}\to 0$ as $n\to\infty$. If
the sequence $\left\{f_{n}\right\}_{n=1}^{\infty}$ is compact,
then $\lambda$ is either a normal eigenvalue, or an eigenvalue
that belongs to the essential spectrum (in the latter case, in
quantum mechanics, such $\lambda$ is called a \textit{bound state embedded
into the continuous spectrum}).

On the Hilbert space 
\[
L_2(T_\infty):=L_2(0,\infty)\oplus \bigoplus\limits_{j=1}^g L_{2}(0,\ell)
\]
of square-integrable vector-valued functions $y=(y_j)_{j=0}^g$ we
introduce an operator $B$, related to the boundary value problem
\eqref{4.1}--\eqref{4.2} with the standard conditions at all vertices, that acts as
\[
B(y_j)_{j=0}^{g}=(-y^{\prime\prime}_{j}+q_{j}y_{j})_{j=0}^g,
\] 
(we recall that $q_0(x)\equiv 0$) with the domain


 \begin{equation} 
 \label{4.3}
D(B):=\left\{y=(y_e)_{j=0}^g \in C(T_\infty)\cap L^2(T_\infty): y''\in L^2(T_\infty) \ \rm{ and }
\right.
\end{equation}
\[\left.
\sum_{e\in E^-_v}y'(0)=\sum_{e\in E^+_v}y'(\ell)\hbox{ for all }v\in V \right\}.
\]

We identify the spectrum of the operator $B$ with the spectrum of the boundary problem \eqref{4.1}--\eqref{4.2} with the standard conditions at all vertices. 

\begin{theorem} (Theorem 3.1 in \cite{MuP}).
Let $q_0\equiv 0$.

 Then  
the operator $B$ on $L_2(T_\infty)$ is self-adjoint and
bounded from below.
Furthermore, $\sigma_{\mathrm{ess}}(B)=[0,\infty)$.
\end{theorem}

Arguments similar to those used in proof of \cite[Theorem 2.1]{LawP} show that the 
restriction
 of the solution of problem \eqref{4.1}--\eqref{4.2} onto the edge $e_0$ is
\begin{equation}
\label{*}
y_0(\lambda,x)=\phi_N(\lambda)\check{s}(\lambda,x)+\phi_D(\lambda)\check{c}(\lambda,x)
\end{equation}
here and in what follows `check' under a letter means correspondence to the case of $q_j\equiv 0$ for all edges. 

Equation (\ref{*}) can be rewritten as
\begin{equation}
\label{4.4}
y_0(\lambda,x)=\frac{1}{2i\sqrt{\lambda}}
\left(e^{i\sqrt{\lambda}x}(\phi_N(\lambda)+i\sqrt{\lambda}\phi_D(\lambda))
-e^{-i\sqrt{\lambda}x}(\phi_N(\lambda)-i\sqrt{\lambda}\phi_D(\lambda))\right).
\end{equation} 
By analogy with the classical $S$-function (see, e.g., \cite{Mar}) we introduce the meromorphic function
\[
S:\lambda\mapsto \frac{E(\sqrt{\lambda})}{E(-\sqrt{\lambda})},
\]
where
\begin{equation}
\label{4.5}
E(\sqrt{\lambda}):=\phi_N(\lambda)+i\sqrt{\lambda}\phi_D(\lambda),\qquad \lambda\in\C,
\end{equation}
is the Jost function. It is cear that $E(\sqrt{\lambda})$ is an entire function of $\sqrt{\lambda}$.

{\bf Theorem 4.2} {\it 1. The zeros of $E(\sqrt{\lambda})$ are located in the closed upper half-plane and on a finite interval of the negative imaginary half-axis. 

2. The number (counting with the multipicities) of the zeros of $E(\sqrt{\lambda})$  on the negative half-axis is the same as the number (counting with the multiplicities) of negative  zeros of the function $\phi_N(\lambda)$.
}

The proof of this theorem can be found in the Appendix. 

\vspace{3mm}

It is known (see, e.g., \cite[Lemma 3.4.2]{Mar}) that 
\[
s_j(\lambda,\ell)\stackrel{|\lambda|\to \infty}{=}\frac{\sin\sqrt{\lambda}\ell}{\sqrt{\lambda}} +O\left(\frac{e^{|{\rm Im}\sqrt{\lambda}\ell|}}{|\lambda|}\right), \quad
c_j(\lambda,\ell)\stackrel{|\lambda|\to \infty}{=}\cos\sqrt{\lambda}\ell +O\left(\frac{e^{|{\rm Im}\sqrt{\lambda}\ell|}}{|\sqrt{\lambda}|}\right).
\]
Substituting these expressions into \eqref{3.8} and \eqref{3.12} we arrive at 

{\bf Lemma 4.3}
{\it Let $q_0=0$.
Then the following asymptotics hold:
\begin{equation}
\label{4.6}
\phi_N(\lambda)\stackrel{\lambda\to +\infty}{=}\check{\phi}_N(\lambda)+o(1), \
\phi_D(\lambda)\stackrel{\lambda\to +\infty}{=}\check{\phi}_D(\lambda)+o(1)
\end{equation}
where `check' in $\check{\phi}_N$ and $\check{\phi}_D$ corresponds to the case of identically zero potentials on all the edges.}

 We will in the following refer to $\check{\phi}_N,\check{\phi}_D$ as the \emph{leading term} of the characteristic function $\phi_N,\phi_D$, respectively.

Substituting \eqref{4.6} into \eqref{4.5} we immediately obtain

{\bf Corollary 4.4}
Let $q_0=0$. Then the asymptotics
\begin{equation}
\label{4.7}
S({\lambda})\stackrel{{\lambda\to +\infty}}{=}\check{S}(\sqrt{\lambda})(1+o(1)),
\end{equation}
and
\begin{equation}
\label{4.8}
E(\sqrt{\lambda})\stackrel{{\lambda\to +\infty}}{=}\check{E}(\sqrt{\lambda})(1+o(1)).
\end{equation}
hold, where quantities with `check' correspond to the case of identically zero potentials on all edges.

\section{ Recovering the shape of a tree by scattering data}

Now we assume that a lead (an half-infinite edge) is attached at one of the vertices of our equilateral  compact tree. Like in \cite{MuP} we assume that the potential $q_0$ is identically zero on the lead while the potentials on the finite edges can be different but all are real and belong to $L_2(0,l)$. Then using the result of \cite{MuP} we  can write that the S-function of our problem on $T_{\infty}$
\[
S(\lambda)=\frac{\phi_N(\lambda)+i\sqrt{\lambda}\phi_D(\lambda)}{\phi_N(\lambda)-i\sqrt{\lambda}\phi_D(\lambda)}.
\]

The  S-function can be found from scattering experiments. Due to \eqref{4.7} knowing the S-function we can use it to find 
\[
\check{S}(\sqrt{\lambda})=\frac{{\check \phi}_N(\lambda)+i\sqrt{\lambda}{\check\phi}_D(\lambda)}{\check{\phi}_N(\lambda)-i\sqrt{\lambda}{\check\phi}_D(\lambda)}.
\]
Since the numerator and the denominator the last fraction are entire functions we can find all the zeros of $\check{E}(\sqrt{\lambda})$ which are not zeros of $\check{E}(-\sqrt{\lambda})$. Now suppose
$\check{E}(\sqrt{\lambda_i})=\check{E}(-\sqrt{\lambda_i})=0$ where $\lambda_i\not=0$.  Then by (\ref{4.5}) we arrive at
$\check{\phi}_N(\lambda_i)=\check{\phi}_D(\lambda_i)=0$ and therefore, $\lambda_i$ is real. Then $\lambda_i$ is an eigenvalue of problem (\ref{3.1})--(\ref{3.6}) or (\ref{3.1})--(\ref{3.4}), (\ref{3.7}) and of problem (\ref{3.1})--(\ref{3.4}), (\ref{3.9}) with $q_j\equiv 0$ for all $j$. Then this $\lambda_i$ is also an eigenvalue of the operator $B$ (the projection of the corresponding eigenfunction onto the lead $e_0$ is identically zero).
 Clearly, in case of identically zero potentials on all the edges the operator $B$ is non-negative and, therefore, the normal eigenvalues does not exist. The embedded eigenvalues may exist on the non-negative axis of $\lambda$-plane. 

  Now let $\check{E}(0)=0$. Then equations (\ref{4.1}) and (\ref{4.2}) take the form $-y''_i=0$ ($i=0, 1, ..., g$). 
Thus, if $0$ belonged to $\sigma (B)$ than $y_0(0,x)$ must be zero almost everywhere on $e_0$. Consequently, $y_j(0,x)\mathop{=}\limits^{a.e.}0$ for all $j$, a contradiction. Thus, $\check {E}(0)\not=0$ and $\check{\phi}_N(0)\not=0$. So, $0$ is not an eigenvalue of $B$ in this case. 

If we know not only S-function but also  all the eigenvalues and their multiplicities then we can find all the zeros of $\check{E}(\sqrt{\lambda})$. Since $\check{E}(\sqrt{\lambda})$ is  a sine-type function we know it up to a constant factor. Using  the representation 
\[
\check{E}(\sqrt{\lambda})=\check{\phi}_N(\lambda)+i\sqrt{\lambda}\check{\phi}_D(\lambda)
\]
we find
\[
\check{\phi}_N(\lambda)=\frac{\check{E}(\sqrt{\lambda})+\check{E}(-\sqrt{\lambda})}{2}, \ \ \ \check{\phi}_D(\lambda)=\frac{\check{E}(\sqrt{\lambda})-\check{E}(-\sqrt{\lambda})}{2i\sqrt{\lambda}}.
\]

Now using (\ref{3.8}) we obtain
\[
\check{\phi}_N(\lambda)=\frac{1}{\sqrt{\lambda}\sin\sqrt{\lambda}l}\psi(\cos\sqrt{\lambda}l)
\]
or

\begin{equation}
\label{5.1}
\psi(z)=\frac{arccos z}{l}\sqrt{1-z^2}\check{\phi}_N\left(\left(\frac{\arccos z}{l}\right)^2\right).
\end{equation}
Using (\ref{3.11})--(\ref{**}) we obtain
\[
\check{\phi}_{D,i}(\lambda)=\hat{\psi}_i(\cos\sqrt{\lambda}l)
\]
\[
\check{\phi}_D(\lambda)=\prod_{i=1}^{d(v_0)}\check{\phi}_{D,i}(\lambda)=\prod_{i=1}^{d(v_0)}det(cos\sqrt{\lambda}l\hat{D}_{T,i}-\hat{A_i})
\]
 \[
\check{\phi}(\lambda)=\prod_{i=1}^{d(v_0)}\hat{\psi}_{i}(\cos\sqrt{\lambda}l)=\hat{\psi}(\cos\sqrt{\lambda}l). 
\]
The last equation gives us
\begin{equation}
\label{5.2}
\hat{\psi}(z)=\check{\phi}_D\left(\left(\frac{arccos z}{l}\right)^2\right).
\end{equation}
Using  (\ref{5.1}) and (\ref{5.2}) we obtain

\begin{equation}
\label{5.3}
\frac{\psi(z)}{\hat{\psi}(z)}=\frac{arccos z}{l}\sqrt{1-z^2}\frac{\check{\phi}_N\left(\left(\frac{\arccos z}{l}\right)^2\right)}{\check{\phi}_D\left(\left(\frac{arccos z}{l}\right)^2\right)}.
\end{equation}
Now we expand the fraction $\frac{\psi(z)}{\hat{\psi}(z)}$ as in Theorem 2.1 and find the shape of the tree.

{\bf Remark 5.1} 1. The common factors of the form $(z^2-\alpha_i)$ with  $\alpha_i\geq 0$  in the numerator and in the denominator of (\ref{5.3}) cancel. These $\alpha_i$ are the eigenvalues of the operator $B$ with $q_j\equiv 0$. They are embedded into the continuous spectrum. Thus, to recover the shape of the tree  we don't need information on the eigenvalues of the initial problem.

2. The length $l$ of an edge can be found taking into account that the functions $\check{\phi}_N(\lambda) $ and $\check{\phi}(\lambda)$ are periodic with the period $\frac{2\pi}{l}$.


\section{Examples}

1. Let 
\[
\check{S}(\lambda)=
\]
\[
\frac{
-\sin \sqrt{\lambda}l\cos\sqrt{\lambda}l(-18\cos^2\sqrt{\lambda}l+13)+i(12\cos^4\sqrt{\lambda}l-17\cos^2\sqrt{\lambda}l+6)
}
{
-\sin \sqrt{\lambda}l\cos\sqrt{\lambda}l(-18\cos^2\sqrt{\lambda}l+13)-i(12\cos^4\sqrt{\lambda}l-17\cos^2\sqrt{\lambda}l+6)
}
\]
Then this function can be presented in the form
\[
\check{S}(\lambda)=-\frac{e(\sqrt{\lambda}l)}{e(-\sqrt{\lambda}l)}
\]
where
\[e(\sqrt{\lambda})=
-\sqrt{\lambda}\cos\sqrt{\lambda}l\frac{(1-\cos^2\sqrt{\lambda}l)}{\sin\sqrt{\lambda}l}(-18\cos^2\sqrt{\lambda}l+13)+
\]
\[
i\sqrt{\lambda}(12\cos^4\sqrt{\lambda}l-17\cos^2\sqrt{\lambda}l+6).
\]
Then using (\ref{5.2}) and (\ref{5.3}) we obtain
\[
\frac{\psi(z)}{\hat{\psi}(z)}=\frac{-36z^5+62z^3-26z}{12z^4-17z^2+6}
\]
or
\[
\frac{\psi(z)}{\hat{\psi}(z)}=-3z-\frac{2}{-3z+\frac{2}{z}}-\frac{1}{-4z+\frac{3}{z}}.
\]
Thus, we arrive at the  form of the tree shown at Fig. 2. By Theorem 2.2 this tree is unique.
\begin{figure}
\begin{center}
   \includegraphics [scale= 0.7 ] {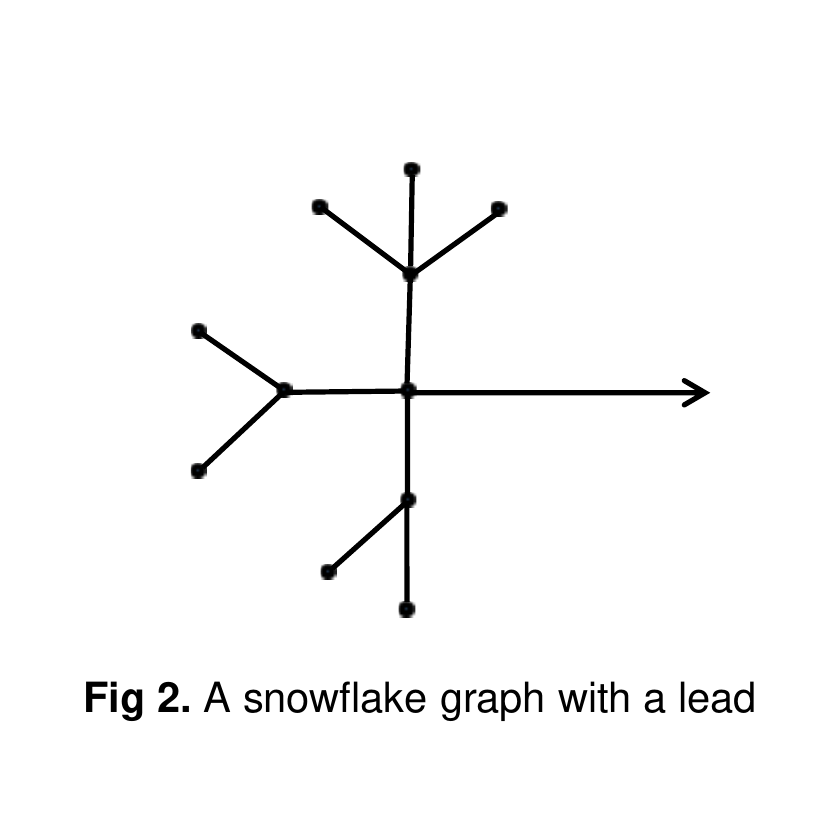}
\end{center}
\end{figure}

2.  Let 
\begin{equation}
\label{6.0}
\check{S}(\sqrt{\lambda})=-\frac{e(\sqrt{\lambda})}{e(-\sqrt{\lambda})}
\end{equation}
where
\begin{equation}
\label{6.00}
e(\sqrt{\lambda})=
\sqrt{\lambda}\sin\sqrt{\lambda}l\cos\sqrt{\lambda}l (9\cos^4\sqrt{\lambda}l-9\cos^2\sqrt{\lambda}l+2)+
\end{equation}
\[
i\sqrt{\lambda}(36\cos^6\sqrt{\lambda}l-60\cos^4\sqrt{\lambda}l+29\cos^2\sqrt{\lambda}l-4)
\]
Then
\[
\frac{\psi(z)}{\hat{\psi}(z)}=\frac{4z(1-z^2)(9z^4-9z^2+2)}{36z^6-60z^4+29z^2-4}=
-z+\frac{1}{3z+\frac{-15z^4+17z^2-4}{12z^5-15z^3+4z}}
\]
The only (up to a permutation) solution  in positive integers of the Diophantine equation 
\begin{equation}
\label{6.1}
\frac{1}{n_1}+ \frac{1}{n_2}=\frac{5}{4}
\end{equation}
is $n_1=1$, $n_2=4$. Therefore,  
\[
\frac{\psi(z)}{\hat{\psi}(z)}=-z+\frac{1}{3z-\frac{1}{z}-\frac{3z^3-2z}{12z^4-15z^2+4}}=
z+\frac{1}{3z-\frac{1}{z}-\frac{1}{4z-\frac{7z^2-4}{3z^3-2z}}}
\]
Since the only (up to the permutation) solution  in positive integers of the Diophantine equation 
\begin{equation}
\label{6.2}
\frac{1}{n_1}+ \frac{1}{n_2}+\frac{1}{n_3}=\frac{7}{3}
\end{equation}
is $n_1=n_2=1$, $n_3=3$ we arrive at
\[
\frac{\psi(z)}{\hat{\psi}(z)}=
z+\frac{1}{3z-\frac{1}{z}-\frac{1}{4z-\frac{2}{z}-\frac{z}{-3z^2-2}}}=z+\frac{1}{3z-\frac{1}{z}-\frac{1}{4z-\frac{2}{z}-\frac{1}{-3z-\frac{2}{z}}}}
\]
According to it the tree must be of the form shown at Fig. 3. Since each of the Diophantaine equations (\ref{6.1}) and (\ref{6.2}) possesses unique solution we conclude that the graph of Fig. 3 is the unique corresponding to the S-function of (\ref{6.0}), (\ref{6.00}).
\begin{figure}
\begin{center}
   \includegraphics [scale= 0.7 ] {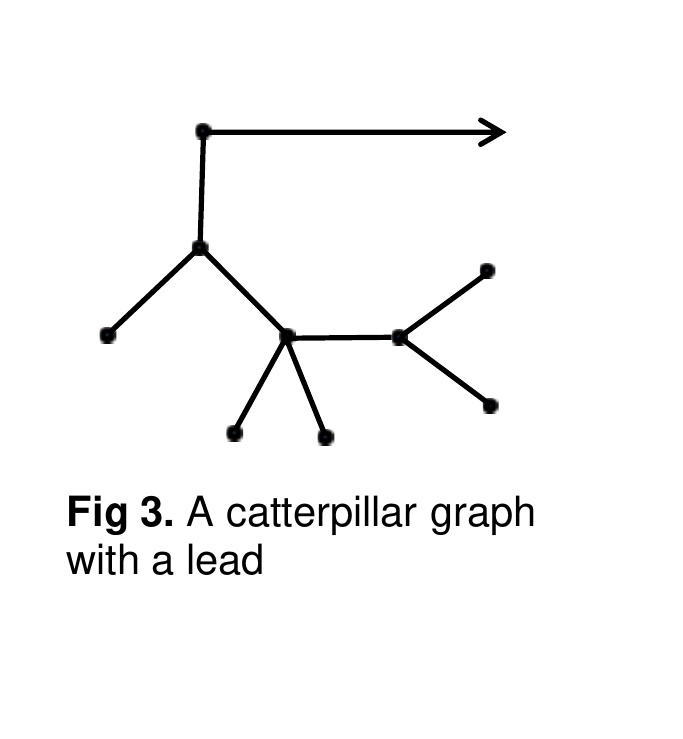}
\end{center}
\end{figure}

\section{Appendix}

{\bf Proof of Theorem 4.2}. Let us consider the following spectral problem
\begin{equation}
\label{7.1}
-y_j^{\prime\prime}+q_j(x)y_j=\lambda y_j, \ \  j=1,2,..., g 
\end{equation} 
where $q_j\in L_2(0,l)$ are real.

For each edge $e_j$ incident with a pendant vertex which is not the root 
we impose the Neumann condition 
\begin{equation}
\label{7.2}
y_j'(l)=0. 
\end{equation}
At  each interior vertex which is not the root we impose the continuity conditions   
\begin{equation}
\label{7.3}
y_j(l)=y_k(0)
\end{equation}
for the incoming to $v_i$ edge $e_j$ and for all $e_k$ outgoing from $v_i$ and the Kirchhoff's conditions
\begin{equation}
\label{7.4}
y'_j(l)=\mathop{\sum}\limits_k y_k'(0)
\end{equation}
where the sum is taken over all edges $e_k$ outgoing from $v_i$. 

If the root is an interior vertex then the conditions at $v_0$ are
\begin{equation}
\label{7.5}
y_i(0)=y_j(0)
\end{equation}
for all indices $i$ and $j$ of the edges incident with the root and
\begin{equation}
\label{7.6}
\mathop{\sum}\limits_i y_i'(0)=i\sqrt{\lambda}y_1(0)
\end{equation}
where the sum is taken over all edges incident with the root ($e_1$ is one of such edges).
If the root is pendant then
\begin{equation}
\label{7.7}
y_1'(0)=i\sqrt{\lambda}y_1(0).
\end{equation}
Consider the operator pencil
\[
\lambda M-i\sqrt{\lambda}K+\tilde{B}
\]

Consider the operator $\tilde{B}$ acting in $L_{2}(0,l)\bigoplus ...
\bigoplus L_{2}(0,l)\bigoplus \C$:
\[
\tilde{B}\left(\begin{array}{c}y_{1}(x) \\ . \\ . \\ . \\ y_{g}(x) \\ c
\end{array} \right)=\left(\begin{array}{c}-y_{1}\pp (x)+q_{1}(x)y_{1}(x) \\ . \\ . \\ . \\
-y_{g}\pp (x)+q_{g}(x)y_{2}(x) \\ y_{1}\p (0)
+...y\p_{d(v_0)}(0) \end{array} \right),
\]

\[
D(\tilde{B})=\left\{\left(\begin{array}{c}y_{1}(x) \\ . \\ . \\ . \\
y_{g}(x) \\ c \end{array}
\right):
\begin{array}{c}y_{j}(x)\in W_{2}^{2}(0,l), \ \ y\p_{j}(0)=0 \ \
for \ \ (j=d(v_0)+1,. . . g),
\\ 
standard \ conditions \ at \ all \ interior \ vertices 
\ except \ of \ the  \ root
\\  y_{1}(0)=. . . =y_{d(v_0)}(0)=c \end{array}\right\}.
\]
\[ 
M=diag\{1,..., 1, 0\}, \ \ K=diag\{0,...,0,1\}.
\]

Introducing the new spectral parameter $\tau=\sqrt{\lambda}$ we obtain quadratic operator pencil we apply Lemma 1.2.1 and Theorem 1.3.3 from \cite{MP0}


\begin{thebibliography}{99}
\itemsep=0pt
%
\bibitem{vB} {J.\ von} Below.
\newblock A characteristic equation associated with an eigenvalue problem on
 $c^2$-networks.
\newblock {\em Lin.\ Algebra Appl.} (1985) 71:309--325.
%
\bibitem{BSS} R. Band, A. Sawicki, U. Smilansky, Scattering from from isospectral quantum graphs. J. Phys. A: Math. Theor. Vol.43 (2010), no. 41.  
%
\bibitem{BSS1} R. Band, A. Sawicki, U. Smilansky, Note on the role of symmetry in scattering from  isospectral graphs and drums. 
math-ph
> arXiv:1110.2475
%
\bibitem{CaP} R. Carlson, V. Pivovarchik. Spectral asymptotics for quantum graphs with equal edge lengths. J. Phys. A: Math. Theor. (2008) 41:145202.
%
%
\bibitem{CP} A. Chernyshenko, V. Pivovarchik. Recovering the shape of a quantum graph. Integr. Equ. Oper. Theory  (2020) 92:23. 
%
%
\bibitem{EE} D.E. Edmunds, W.D. Evans. Spectral theory and differential operators. Clarendon Press, Oxford, 1989.
%
\bibitem{GK} I. Gohberg, M. Krein. Introduction to the Theory of Linear Non-Selfadjoint Operators in Hilbert Space. AMS, 1969.
%
\bibitem{GS} B. Gutkin, U. Smilansky, Can one hear the shape of a graph? J. Phys. A Math. Gen. (2001), 34:6061--6068. 
%
\bibitem{HLBSKS} O. Hul, M. Lawniczak, S. Bauch, A. Sawicki, M. Kus, L, Sirko. Are Scattering Properties of Graphs Uniquely Connected to Their Shapes?
Phys. Rev. Lett. Vol. 109 (2012), 040402.
%
\bibitem{KurSte02}
P.\ Kurasov and F.~Stenberg.
\newblock {On the inverse scattering problem on branching graphs}.
\newblock {\em J.\ Phys.\ A} (2002) 35:101--121.
%
\bibitem{LP} Y. Latushkin, V. Pivovarchik. Scattering in a forked-shaped waveguide. Integr. Equ. Oper. Theory (2008) 61:365--399. 
%
\bibitem{LawP} C.-K. Law, V. Pivovarchik. Characteristic functions of quantum graphs. J. Phys A: Math. Theor. (2009) 42:035302.
%
\bibitem{Mar} V.A. Marchenko. Sturm--Liouville Operators and Applications, revised edition. AMS Chelsea Publishing, Providence, RI 2011. 
%
%
\bibitem{MP0} M. M\" oller, V. Pivovarchik, Spectral Theory of Operator Pencils, Hermite--Biehler Functions, and Their Applications. Birkhäuser, Cham, 2015.
%
%
\bibitem{MP}  M. M\" oller, V. Pivovarchik, Direct and inverse finite-dimensional spectral problems on graphs
Operator Theory: Advances and Applications, 283. Birkh\"auser/Springer,  2020. 359 pp. ISBN: 978-3-030-60483-7; 978-3-030-60484-4 https://www.springer.com/gp/book/9783030604837
%
\bibitem{MuP} D. Mugnolo, V. Pivovarchik. Distinguishing co-spectral quantum graphs by scattering, J. Phys. A: Math. Theor., Vol. 56, issue 9, (2023) DOI: 10.1088/1751-8121/acbb44, 
  arXiv: 2211.05465. 
%
\bibitem{OSh} Y. Okada, A. Shudo,  S. Tasaki, T. Harayama. Can one hear the shape of a drum?: revisited. J. Phys. A: Math. Gen. ,  Vol. 38 (2005), L 163.
%
\bibitem{P} V. Pivovarchik. Recovering the shape of a quantum treeby two spectra. ArXiv: 2301.05939. Submitted to Inverse Problems and Imaging.
\bibitem{P1} V. Pivovarchik. Scattering in a loop-shaped waveguide. In: Recent Advances in Operator Theory, Groningen, 1998. Birkhäuser, Basel and Boston, 2001: 527--543. 
%
\bibitem{Re} T. Regge. Construction of potential from resonances. Nuovo
Cimento  (1958) 9:491--503 and 671--679.
%




\end{thebibliography}
\end{document}